\documentclass[aps,pre,preprint,showpacs,amsmath,amssymb,superscriptaddress]{revtex4}

\usepackage{graphicx}

\usepackage{graphicx}
\usepackage{amsmath,amssymb,latexsym}
\usepackage{color}
\bibstyle{apsrev.bib}

\begin{document}
\def\be{\begin{equation}}
\def\ee{\end{equation}}

\def\bfi{\begin{figure}[h]}
\def\efi{\end{figure}}
\def\bea{\begin{eqnarray}}
\def\eea{\end{eqnarray}}

\title{Effects of Frustration on Fluctuation-dissipation Relations}

\author{Federico Corberi}
\email{corberi@sa.infn.it}
\affiliation {Dipartimento di Fisica ``E.~R. Caianiello'', and INFN, Gruppo Collegato di Salerno, and CNISM, Unit\`a di Salerno,Universit\`a  di Salerno, 
via Giovanni Paolo II 132, 84084 Fisciano (SA), Italy.}

\author{Manoj Kumar}
\email{manojkmr8788@gmail.com}
\affiliation{International Centre for Theoretical Sciences, Tata Institute of Fundamental Research, Bengaluru 560089, India.}

\author{Eugenio Lippiello}
\email{eugenio.lippiello@unina2.it}
\affiliation{Department of Mathematics and Physics, University of Campania ÒL. VanvitelliÓ, Viale Lincoln 5, 81100 Caserta, Italy.}

\author{Sanjay Puri}
\email{purijnu@gmail.com}
\affiliation{School of Physical Sciences, Jawaharlal Nehru University, New Delhi 110067, India.}

\date{\today}

\begin{abstract}

We study numerically the aging properties of the two-dimensional Ising model with quenched disorder considered in our recent paper [Phys. Rev. E {\bf 95}, 062136 (2017)], where frustration can be tuned by varying the fraction a of antiferromagnetic interactions. Specifically we focus on the scaling properties of 
the autocorrelation and linear response functions after a quench of the model to a low
temperature. We find that the interplay between equilibrium 
and aging occurs differently in the various regions of the phase diagram of the model. When the quench is made into the ferromagnetic phase
the two-time quantities are made by the sum of an equilibrium and an aging part, 
whereas in the paramagnetic phase these parts combine in a multiplicative way. Scaling forms are shown to be obeyed with good accuracy, and
the corresponding exponents and scaling functions are determined and discussed in the framework of what
is known in clean and disordered systems. 
   
\end{abstract}
\maketitle

\section{Introduction}

After a quench from a high temperature to a low-temperature phase a system enters a dynamical state which is generally characterized
by slow evolution and aging. In the simplest cases, as in binary systems without quenched disorder, 
the kinetics is quite well understood. Domains of the two low-temperature equilibrium phases form and evolve
with an average size $L(t)\propto t^{1/z}$ growing algebraically as time elapses. A prominent feature is the existence
of dynamical scaling, namely the fact that configurations of the same system at different times happen to be self similar, namely
statistically equal except for a trivial rescaling of lengths by a factor $L(t)$. As a consequence, observable quantities such as
correlation functions and linear response functions  take a definite scaling form, 
similarly to what happens in static critical phenomena.

The value of the dynamical exponent $z$ and of other exponents entering such scaling forms, together with the behavior of the
scaling functions, are known to exhibit universal properties, being dependent only on few relevant features, e.g. the scalar or
vectorial nature of the system, and the presence of conservation laws or hydrodynamic interactions.  
In addition, the connection between static and dynamic properties derived in~\cite{franz98} 
allows one to infer the form of the so called fluctuation-dissipation plot, namely the asymptotic relation between two two-time
quantities, the autocorrelation function and the associated linear response
function, starting from the well known structure of the broken-symmetry equilibrium state.
 
Such a good comprehension of the universal properties of the kinetics is, however, lost as soon as
quenched disorder is present in the system. This is true not only in the case when strong disorder and frustration are present,
such as in the emblematic case of spin-glasses, but even for tiny amounts of quenched randomness.
Indeed, even in the cases when a weak disorder does not produce relevant changes in the static properties,
the non-equilibrium kinetics is usually affected in a dramatic way. In magnets, for instance, the presence of unequal,
though ferromagnetic, coupling constants may change the exponent of the power-law growth of 
$L(t)$ with respect to the clean case or even turn $L(t)$ into a logarithmic form~\cite{corberi2011b}.
Similar features are observed in the presence of other kinds of quenched randomness~\cite{corberi2012,corberi2013,lippiello2010,Castellano1998}.

When disorder is stronger and associated with frustration the problem is by far much more complicated 
and the interpretation of both the static structure and the non-equilibrium properties are still debated issues~\cite{corberi2011}. 
In this scenario, understanding the properties of two-time quantities and of their relation might represent an important contribution
to the understanding of both the static and the dynamics, given the bridge between these two aspects 
provided by the fluctuation-dissipation relation discussed above. However, given the very slow and complex evolution affecting these
systems, arriving to a conclusive numerical evidence on the asymptotic form of correlation and response functions 
often represents a formidable task.

In this paper we tackle this matter from a different perspective. Instead of facing the hard problem straightforwardly, by measuring 
two-time quantities
directly in the fully frustrated system, we try to narrow the difficulty, starting from the well understood clean ferromagnetic case and moving 
towards the fully frustrated case in a model where the amount of disorder and frustration can be tuned at will.
We do that by computing
numerically the spin autocorrelation function
$C(t,t_w)=N^{-1}\sum _{i=1}^N\langle s_i(t)s_i(t_w)\rangle$
and the associated linear response in a random-bond Ising model with an adjustable fraction $a$ of 
antiferromagnetic bonds. In this way we can study the
modifications of the properties of two-time quantities when, starting from the well known clean ferromagnet with $a=0$, one moves
into a phase where frustration is relevant. 

It is important to highlight the novel and important features in our modeling and results presented in this paper. The first significant aspect we address is the universality of two-time quantities. According to the {\it superuniversality} hypothesis~\cite{cugliandolo2010, fisher1988}, once expressed in a scaling form in terms of the growing length $L(t)$, exponents and scaling functions of different quantities in a coarsening system are independent of the nature and magnitude of quenched disorder. We find that this is not the case in the present model. In particular, the {\it response function exponent}, which has been the subject of many recent studies~\cite{barrat1998,corberi2001,corberi2001b,henkel2001,henkel2003,henkel2003b,corberi2003,corberi2003b,corberi2004b,corberi2005,corberi2005b,lippiello2005,henkel2005,lippiello2006}, turns out to be strongly disorder-dependent. Our results show that it vanishes at the transition from the ferromagnetic to the paramagnetic phase.

Our model does not exhibit a spin-glass phase in the $d=2$ case considered in this paper. Notwithstanding, 
spin-glass order is expected at zero temperature when frustration is strong enough. The $d=2$ case has the advantage that low-temperature equilibrium states can be determined relatively rapidly, thus allowing us to consider systems with sufficiently large sizes. A precise determination and understanding of the two-time quantities is therefore possible, both in the equilibrium states and in the non-equilibrium evolution. This is the second important aspect of our present study. In this context, we use an out-of-equilibrium generalization of the fluctuation-dissipation theorem \cite{lippiello2005,lippiello2008,lippiello2008b,baiesi2009,corberi2010}. Thus, we are able to cleanly address the issue of how the equilibrium and the aging degrees of freedom combine to determine the scaling forms of correlation and response functions. This is an important pre-requisite to understand the properties of the fluctuation-dissipation relation~\cite{lippiello2005,lippiello2008,lippiello2008b,baiesi2009,corberi2010}. To the best of our knowledge, this approach has never been pursued for this kind of model. We find that the structure of the phase diagram (see Fig.~\ref{fig_phase_diagr}) is faithfully reflected in the properties of the two-time quantities. The phase diagram shows a ferromagnetic phase for $a<a_f$, and an antiferromagnetic phase for $a>a_a$. These are separated by a paramagnetic region (with spin-glass order at $T=0$) for $a_f\le a \le a_a$. The corresponding properties of the two-time quantities are as follows: \\
(a) In the ferromagnetic phase ($a<a_f$) (and similarly in the antiferromagnetic phase with $a>a_a$), the two-time quantities have an additive form. For example, the autocorrelation function obeys $C(t,t_w)=C_{eq}(t-t_w)+C_{ag}(t,t_w)$. Here, $C_{eq}$ is the equilibrium correlation, and the aging part $C_{ag}$ obeys a scaling form $C_{ag}(t,t_w)=c[L(t)/L(t_w)]$ ($c$ is a scaling function). (An analogous behavior is shown by the response function.) This structure is generic, but the scaling functions and the response function exponent depend on $a$, with the latter vanishing at $a=a_f$. The fluctuation-dissipation plot has the usual broken-line shape of ferromagnetic models~\cite{corberi2001b}. \\
(b) At the critical point $a=a_f$ (or, equivalently, $a=a_a$), the additive structure turns into a multiplicative one   $C(t,t_w)=C_{eq}(t-t_w)\cdot C_{ag}(t,t_w)$, with the scaling properties of $C_{ag}$ as discussed above. The fluctuation-dissipation plot also changes radically, and now approaches the equilibrium linear behavior expected in systems at  criticality \cite{godreche2002,calabrese2005}. These properties are observed in the whole 
frustration-dominated paramagnetic region $a_f\le a\le a_a$.

This paper is organized as follows: In Sec.~\ref{model} we introduce and describe the model and the quantities that will be
numerically computed. In Sec.~\ref{scaling} we briefly review what is known about the scaling properties of two-time quantities
in clean and disordered systems. In Sec.~\ref{numresults} we present and discuss the outcomes of our numerical simulations.
Sec.~\ref{concl} concludes the paper with a summary and a discussion of our findings and of possible future perspectives. 

\section{Model and observable quantities} \label{model}

In this paper we consider the spin model governed by the Hamiltonian
\be
{\cal H}(\{s_i\})=-\sum _{\langle ij\rangle}J_{ij}s_is_j,
\label{ham}
\ee
where $s_i = \pm 1$, $\langle ij\rangle$ denotes nearest neighbors sites of a lattice, and 
$J_{ij} = J_0 + \xi _{ij}$ are uncorrelated stochastic random couplings with $J_0 > 0$ and $\xi _{ij}$ 
extracted from a bimodal distribution
\be
P(\xi)=a\,\delta _{\xi ,-K} +(1-a)\,\delta _{\xi,K}, 
\ee
where $K>J_0$, $0\le a\le 1$ is the fraction of antiferromagnetic bonds, and $\delta $ is a Kronecker function.
We will consider a square lattice with $N$ spins, periodic boundary conditions, $J_0=1$ and $K=5/4$ in the following. 
We also set the Boltzmann constant $k_B=1$. With these parameters the model has been previously characterized in~\cite{corberi2017}.
A similar model was introduced and studied in \cite{Ozeki1987,hartmann1999}.

It was shown~\cite{corberi2017} that the model has a ferromagnetic phase for sufficiently low
temperatures $T<T_c(a)$, where $T_c(a)$ is a critical temperature vanishing in the limit $a\to a_f^-$, with $a_f\simeq 0.4$,
as pictorially sketched in Fig.~\ref{fig_phase_diagr} (the reason for using such a rough representation is that the real phase-diagram of this model,
namely a determination of $T_c$ for all values of $a$, is not currently available).
For large values of $a$ an antiferromagnetic phase exists (which will not be considered 
in this paper) for $T<T_c(a)$ (we use the same symbol as for the ferromagnetic case for simplicity), where 
$T_c(a)\to 0$ as $a\to a_a^+$. 
For intermediate values of $a$ 
the equilibrium state of the system is paramagnetic and disordered
except, possibly, right at $T=0$ where for $a_f\le a \le a_a$ a spin-glass phase is expected. 
\bfi
\begin{center}
\includegraphics*[width=0.95\columnwidth]{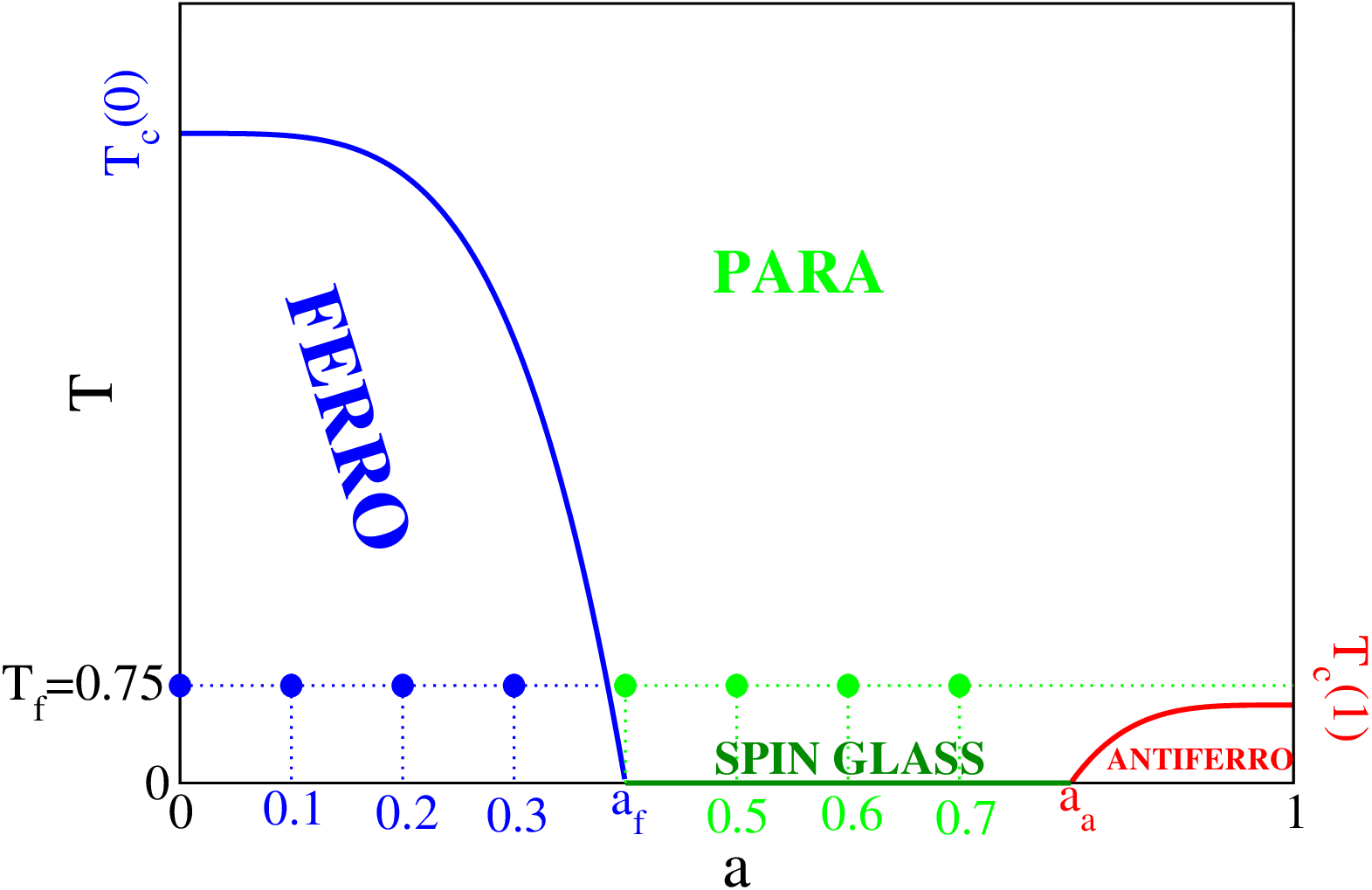}
\end{center}
\caption{Pictorial representation of the phase-diagram of the model. The heavy dots are the points where our simulations are carried out.}
\label{fig_phase_diagr}
\efi

In the following we will consider the non-equilibrium kinetics obtained by quenching the present model from an equilibrium
state at infinite temperature to a sufficiently low final temperature $T_f$. The values of $a$ studied in our simulation,
and the value of $T_f$ considered, are also shown in Fig.~\ref{fig_phase_diagr}.

In order to detect the build-up of spatial correlations
an average size of ordered regions at time $t$ can be defined as
\be
L(t)=[E(t)-E_{eq}]^{-1},
\ee
where $E(t)$ is the energy per spin of the system at the current time $t$ and $E_{eq}$ is the same quantity
computed at equilibrium at the temperature $T=T_f$. This definition of an ordering length is standard
in ferromagnetic systems. A thorough discussion, together with a determination of this quantity, can be found in~\cite{corberi2017}.

In this paper we focus on the scaling properties of the two-time quantities that we detail below.
The autocorrelation function is defined as
\be
C(t,t_w)=\frac{1}{N}\sum _{i=1}^N[ \langle s_i(t)s_i(t_w)\rangle- \langle s_i(t)\rangle \langle s_i(t_w)\rangle],
\label{defC}
\ee
with $t\ge t_w$. Here $\langle \dots \rangle $ means both the thermal average, namely over initial conditions 
and dynamical trajectories, and over the realizations of the quenched disorder. 
Notice that, after a quench from high temperature, symmetry is not broken at any finite
time (in the thermodynamic limit) and hence the subtraction term on the r.h.s. of Eq.~(\ref{defC}) is immaterial
(however it will be relevant when we will introduce the autocorrelation in equilibrium $C_{eq}$, below). 
The impulsive autoresponse function is defined as
\be
R(t,t_w)=\frac{1}{N}\sum _{i=1}^N \left . \frac{\delta \langle s_i(t)\rangle _h}
{\delta h_i(t_w)}\right \vert _{h=0},
\ee
where $h_i(t)$ is a time-dependent magnetic field and $\langle \dots \rangle_h$
means an average in the presence of such field. Since this quantity is very
noisy it is customary to measure the so called {\it integrated}
autoresponse function, sometimes also denoted as the
{\it zero-field cooled susceptibility}
\be
\chi(t,t_w)=\int _{t_w}^t dt'\, R(t,t').
\label{defchi}
\ee
This quantity has an enhanced signal/noise ratio and is more suited to
enlighten the scaling properties, as discussed in~\cite{corberi2003}.
In order to compute $\chi$ numerically without applying the small perturbation we use the generalization of
the fluctuation-dissipation theorem to non-equilibrium states derived
in~\cite{lippiello2005,lippiello2008,lippiello2008b,baiesi2009,corberi2010}.

\section{Scaling behaviors} \label{scaling}

\subsection{Non disordered systems}

The scaling behavior of the two-time quantities introduced in Sec.~\ref{model}
are quite well understood in ferromagnetic models without quenched
disorder and short-range interactions~\cite{corberi2004,corberi2011}. 
In this case slow relaxation is observed
when a system in equilibrium at $T>T_c$ is quenched either to the critical
temperature $T_c$ or to any final temperature $T_f<T_c$, including $T_f=0$.
In any case, after a transient, a dynamical state is entered
where an ordering length grows algebraically in time as
$L(t)\sim t^{1/z}$~\cite{bray94}, where $t=0$ is the quench instant.
Referring to the case of a purely relaxational dynamics with a
non-conserved order parameter considered in the present paper
the exponent $z$ takes a value $z=2$ independent of both
$T_f$ and spatial dimension $d$ in all quenches with $T_f<T_c$.
Instead, when the quench is made at $T_f=T_c$,
$z$ coincides with the dynamical critical exponent $z_c$ which
depends on $d$ and becomes $z_c=2$ only at the lower critical dimension
$d_L$~\cite{hohenberg77}. 

It should be stressed that, although the unbounded
growth of $L(t)$ makes the non-stationary character of the dynamics
manifest, on sufficiently short time/space scales local equilibration takes place.
For instance, in the ferromagnetic systems we are considering now, 
thermal fluctuations well inside the overall ordered growing domains,
whose typical size is the equilibrium coherence length $\xi _{eq}(T_f)\ll L(t)$,
behave as in an equilibrium state at $T=T_f$. As we will see shortly, the 
aging (i.e. non-equilibrium) and equilibrium features
may combine differently in determining the scaling properties of 
the observable quantities.  

The self-similarity of configurations as time elapses determines
a dynamical scaling symmetry which, in turn, informs
observable quantities such as, e.g., $C(t,t_w)$ and $R(t,t_w)$ or $\chi(t,t_w)$.
As discussed in~\cite{corberi2004}, the scaling properties of these functions depend, 
in turn, on the kind of quench. More precisely one has three different
behaviors corresponding to i) a subcritical quench to  $T_f< T_c$, 
ii) a critical quench to $T_f=T_c>0$ for $d>d_L$
and iii) a critical quench to $T_f=T_c=0$ for $d=d_L$.
We discuss them separately below.

\subsubsection{Sub-critical quench to $T_f<T_c$ ($d>d_L$)} \label{additive}

In this case, for large $t_w$, $C$ and $R$ take the forms~\cite{corberi2001b}
\be
C(t,t_w)=C_{eq}(t-t_w)+C_{ag}(t,t_w)
\label{Cadd}
\ee
\be
R(t,t_w)=R_{eq}(t-t_w)+R_{ag}(t,t_w),
\label{Radd}
\ee
where $C_{eq}$ and $C_{ag}$ (and similarly for $R$) are an equilibrium and an aging term, respectively.
The former is the one that one should have in a system in equilibrium at the final temperature of the quench and obeys the fluctuation-dissipation theorem
\be
TR_{eq}(t-t_w)=-dC_{eq}(t-t_w)/dt
\label{fdtR}
\ee
and the latter is what is left over.
Notice that this is an additive structure where equilibrium and non-equilibrium contributions sum up.
The aging parts obey a scaling form
\be
C_{ag}(t,t_w)=c\left ( \frac{L(t)}{L(t_w)}\right )
\label{Cagscalbelow}
\ee
and
\be
R_{ag}(t,t_w)=L(t_w)^{-(z+\alpha)}r\left ( \frac{L(t)}{L(t_w)}\right ),
\label{Rscaladd}
\ee
where $c$ and $r$ are scaling functions 
and $\alpha $ is the
response function exponent.
Notice that Eq.~(\ref{Cagscalbelow}) could also be written as $C_{ag}(t,t_w)=L(t_w)^{-\beta}c \left(\frac{L(t)}{L(t_w)}\right)$
with $\beta=0$ and this implies~\cite{coniglio2000} that the domains grow with a dimension $d-z\beta/2=d$, namely they
are compact for this kind of quench. The exponent $\alpha>0$ is not
related to the behavior of $C$ (at variance with the case of a critical quench, see Sec.~\ref{multiplicative} below), and its determination has been the subject of many studies~\cite{barrat1998,corberi2001,corberi2001b,henkel2001,henkel2003,henkel2003b,corberi2003,corberi2003b,corberi2004b,corberi2005,corberi2005b,lippiello2005,henkel2005,lippiello2006,mazenko2004}, both on the analytical and numerical side. 

It is a trivial consequence of Eqs.~(\ref{Radd},\ref{Rscaladd}) 
that an additive structure informs also the integrated response,
$\chi(t,t_w)=\chi_{eq}(t-t_w)+\chi_{ag}(t,t_w)$, with 
\be
\chi_{ag}(t,t_w)=L(t_w)^{-\alpha}f\left  ( \frac{L(t)}{L(t_w)}\right ).
\label{chiagferro}
\ee
From the properties discussed above we see that in the short time-difference regime one has
\be
C(t,t_w)=C_{eq}(t-t_w)+q_{EA},
\ee
where $q_{EA}=c(1)$ is the so called Edwards-Anderson order parameter which, for a ferromagnet, amounts to
the squared spontaneous magnetization at equilibrium at $T=T_f$.
Instead, in the large time-difference regime, namely with $t-t_w\to \infty$ with fixed $L(t)/L(t_w)$, 
one has
\be
C(t,t_w)=C_{ag}(t,t_w).
\ee
Starting from these behavior it is easy to show~\cite{corberi2004} that $C$ has the weak ergodicity breaking
property $\lim _{t-t_w\to \infty}\lim _{t_w\to \infty (t-t_w fixed)}C(t,t_w)\neq   \lim _{t_w\to \infty}\lim _{t-t_w\to \infty (t_w fixed)}C(t,t_w)$,
which is associated to the broken ergodicity of the equilibrium state below $T_c$. This is at variance to what happens
in the critical quench (see Sec.~\ref{multiplicative}), where spontaneous magnetization does not develop
and ergodicity occurs.  

For the response function one has that $R_{eq}$ obeys the fluctuation-dissipation theorem~(\ref{fdtR})
with respect to $C_{eq}$ and vanishes in the large time-difference regime while, conversely, $R_{ag}$
vanish in the short time-difference regime.

Let us mention that the above features are independent of $T_f$ and hence apply down to $T_f=0$,
since temperature is an irrelevant parameter in the renormalization group sense~\cite{mazenko1985,bray1990}.

\subsubsection{Critical quench to $T_f=T_c>0$ ($d>d_L$)} \label{multiplicative}

In this case the forms~(\ref{Cadd},\ref{Radd}) change to~\cite{godreche2002,calabrese2005}
\be
C(t,t_w)=C_{eq}(t-t_w)C_{ag}(t,t_w)
\label{Cmult}
\ee
\be
R(t,t_w)=R_{eq}(t-t_w)R_{ag}(t,t_w)
\label{Rmult}
\ee
where $C_{ag}$ and $R_{ag}$ are the non-equilibrium contributions which depend only
on the ratio $L(t)/L(t_w)$
\be
C_{ag}(t,t_w)=c\left (\frac{L(t)}{L(t_w)}\right )
\label{scalCcrit}
\ee
\be
R_{ag}(t,t_w)=r\left (\frac{L(t)}{L(t_w)}\right ),
\label{scalRmult}
\ee
with $c(x)$ and $r(x)$ scaling functions (different from the ones of Eqs.~(\ref{Cagscalbelow},\ref{Rscaladd})) ,
 whereas 
\be
C_{eq}(t-t_w)=(t-t_w+t_0)^{-B}
\label{Ceqmult}
\ee
and
\be
R_{eq}(t-t_w)=(t-t_w+t_0)^{-(1+A)},
\ee
are the equilibrium autocorrelation and response functions at $T=T_c$. 
Here $A$ and $B$ are the autocorrelation and response exponents,
$t_0$ is a microscopic time and the fluctuation-dissipation
theorem~(\ref{fdtR})  
implies $A=B$. A scaling relation links the actual value of these exponents 
to the usual equilibrium critical static and dynamic ones $\eta $ and $z$ through $A=B=(d-2+\eta)/z$ which,
in turn, implies~\cite{coniglio2000} a fractal dimension $D=d-zB /2$ of the critical correlated clusters.
Notice that $B\to 0$ as $d\to d_L^+$, implying that critical clusters become compact objects in this limit.

Let us stress that the structure~(\ref{Cmult},\ref{Rmult}) means that the equilibrium part and the non-equilibrium
one of two-time quantities enter in a multiplicative manner. Replacing the form~(\ref{Rmult},\ref{scalRmult}) into 
Eq.~(\ref{defchi}) one finds that no particular scaling property shows up in $\chi(t,t_w)$. However, 
it can be shown~\cite{corberi2006} that the quantity $T_f[1-\chi(t,t_w)]$, which represents the 
distance from the equilibrium static value, scales as
\be
[1-T_f\chi(t,t_w)]=L(t_w)^{-\gamma}g\left (\frac{L(t)}{L(t_w)}\right ),
\label{distance}
\ee   
where $\gamma=zB$ and $g$ is a scaling function.

According to Eqs.~(\ref{Cmult},\ref{Rmult}) in the short time difference regime, namely letting 
$t_w$ become large while keeping $t-t_w$ fixed, one gets
 \be
C(t,t_w)=C_{eq}(t-t_w)C_{ag}\left (1\right )\propto C_{eq}(t-t_w)
\label{Cmultshort}
\ee
\be
R(t,t_w)=R_{eq}(t-t_w)R_{ag}\left (1\right )\propto R_{eq}(t-t_w).
\label{Rmultshort}
\ee

\subsubsection{Quenches to $T_f=0$ with $d=d_L$}

At $d_L$ it is $T_c=0$ and hence $T_f=0$ can be viewed also as a limiting case of a critical quench. However,
since an equilibrium system without quenched disorder is perfectly ordered at $T=0$, it is clear that the scaling structure
of two-time quantities must be akin to the one of the subcritical quenches, namely additive, because 
weak ergodicity breaking must occur. Moreover $C=C_{ag}$,
since the equilibrium state at $T=0$ has no dynamics. The same property is shared by $\chi _{eq}$, but only in scalar 
systems with a discrete (up-down) symmetry, while $\chi _{eq}$ does not vanish in vectorial systems with continuous
symmetry, due to the presence of Goldstone modes.
The distinguishing feature of the quench at $T_f=0$ with $d=d_L$ 
is that $\alpha=0$ in this case~\cite{lippiello2000,godreche2000,corberi2001,corberi2001b,corberi2002,corberi2002b,corberi2002c,
corberi2003,burioni2006,burioni2007}.

\subsection{Disordered systems}

Let us now briefly discuss the modifications due the presence of quenched
disorder (see also~\cite{berthier2002,jaubert2007} for a discussion of this topic).
Since the matter is, in some cases, still debated, we will make only reference to some systems
where a good understanding and accepted analytic background is available. 

It turns out that the presence of quenched disorder may introduce a different scaling pattern with respect to
those encountered in so far.  When systems such as 
$p$-spins~\cite{cugliandolo93} or mean field spin glasses~\cite{cugliandolo94,mezard87} 
are quenched to a phase with $q_{EA}>0$, namely to below a finite critical temperature, one has an additive structure, as 
expected.  However, at variance with the non-disordered case discussed in Sec.~\ref{additive},
one has a value $\alpha =0$ of the response function exponent for $d>d_L$. We remind that $\alpha =0$ is found
also in clean ferromagnetic systems but only at $d_L$, where the quench can only be made at $T_f=0$ and,
due to that, $C_{eq}$ and $\chi _{eq}$ vanish identically in system with a scalar order parameter. 
Conversely, in the aforementioned disordered scalar models, for $d>d_L$
quenches with $T_f>0$ display both $\alpha =0$ and a non-trivial $C_{eq}$ and $\chi _{eq}$.

\section{Numerical results} \label{numresults}

We have run a set of simulations both of a system quenched from infinite temperature to a final temperature
$T_f=0.75$ (we set the Boltzmann constant equal to unity) and, in parallel, of the same system in equilibrium 
at the final temperature of the quench. In order to equilibrate the system we first found the ground state 
by means of the algorithm discussed in~\cite{Khoshbakht2018} which allows to determine the configuration
in polynomial time. Then, using the ground state as an initial condition, we have equilibrated the system at the
working temperature by means of standard Montecarlo techniques. We have simulated two-dimensional
systems on a square lattice with 512 x 512 spins. This size is free from finite-size effects in the accessed time window. For any run we have taken an average over $10^3$ different realizations.
Montecarlo moves use Glauber transition rates.   

The chosen value of $T_f$ has been 
shown to be a reasonable compromise between the attempt to study the low-temperature behavior of the model and
the need to avoid the sluggish dynamics observed when $T_f$ is too low. Our simulations are performed for several 
values of $a$ in order to span both the ferromagnetic and the paramagnetic region (the antiferromagnetic
phase is expected to give similar results to the ferromagnetic one). A visual summary of the various quenches considered
in this study is provided in Fig.~\ref{fig_phase_diagr}.
We also acknowledge that a study of the autocorrelation function in a related model (but restricted to the case $d=3$,
which is rather different due to the presence of a spin-glass phase at finite temperature) was carried out
in~\cite{manssen2015}. To the best of our knowledge, the response function has never been
considered.

The behavior of the ordering length $L(t)$ after a quench of the model has been thoroughly discussed 
in~\cite{corberi2017}. For completeness we show in Fig.~\ref{compare_L} its behavior 
for the values of $a$ that will be considered in this
paper. Here and in the following, time is measured in units of Montecarlo steps.
Notice that in Fig.~\ref{compare_L} we normalize $L(t)$ by its value at an early time ($t=4$)
in order to better compare curves with different $a$. One observes that, in the range of times considered,
$L(t)$ keeps growing for any value of $a$. The growth is faster in the pure case (when one has 
$L(t)\propto t^{1/2}$) than for any other value of $a$, and the slowest case occurs with $a=0.2$.
The fact that $L(t)$ keeps growing also in the paramagnetic phase can be interpreted as due to the proximity of
the spin glass phase at $T=0$, for $a_f<a<a_a$, as will be further discussed below.

\bfi
\begin{center}
\includegraphics*[width=0.95\columnwidth]{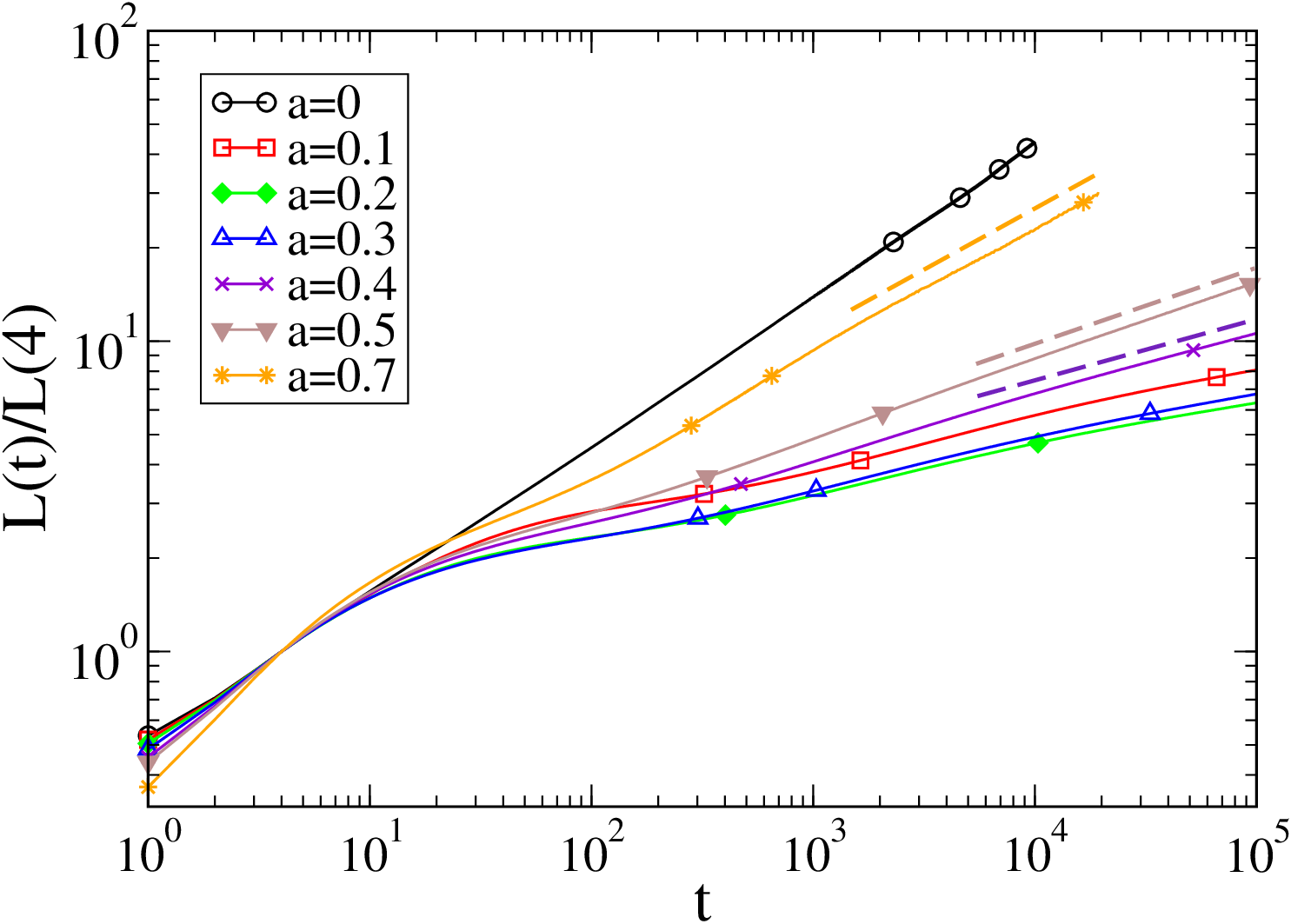}
\end{center}
\caption{$L(t)$ is plotted against time for different values of $a$ (see legend) after a quench of the model to $T_f=0.75$.
The dashed lines slightly above the right part of some of the curves are the best algebraic fits. Specifically the indigo line is the behavior $x^{1/5}$, the dashed brown is the one $x^{1/4}$, and the dashed orange is $x^{1/2.56}$.}
\label{compare_L}
\efi

In the following we will discuss the behavior of the two-time quantities introduced in Sec.~\ref{model},
separating the discussion for the different phases of the model. 

\subsubsection{Quenches with $a<a_f$}

In this section we present data for values of $a<a_f$. We have checked that for all the quenches
studied in this sector one has $T_f\ll T_c(a)$, namely that the target equilibrium state is in the ferromagnetic phase.
According to the discussion of Sec.~\ref{additive}, when the quench is done in a phase
where symmetry breaking occurs and $q_{EA}>0$, one expects an additive structure
for the two-time quantities. For the autocorrelation, according to 
Eqs.~(\ref{Cadd},\ref{Cagscalbelow}), one should find data collapse, for any given value of $a<a_f$, by plotting  
$C(t,t_w)-C_{eq}(t-t_w)$ against $L(t)/L(t_w)$. We have computed $C(t,t_w)$ in the quenched system
and $C_{eq}$ in the equilibrium state and the result of this plot is shown in Fig.~\ref{compare_C} for different
values of $a$. Notice that in this plot we use $L(t)/L(t_w)-1$ on the $x$-axis in order to better show the
small time-difference regime. Fig.~\ref{compare_C} shows an excellent data collapse both for $a=0.1$ and for $a=0.3$
(a similar quality of the collapse is obtained also for other values of $a<a_f$, not shown here).
This proves quite convincingly that the scaling structure described in Sec.~\ref{additive} applies also to
the present disordered case. 

Notice that, for any $a>0$, the scaling function $g$ appearing in Eq.~(\ref{Cagscalbelow}) is markedly
different from the one of the pure case (plotted with a bold green curve). 
It must be kept in mind that this difference is trivially due, at least in part, to the different value of $q_{EA}$ as 
$a$ changes. However this fact is not sufficient to explain the differences between the curves for different $a$.
To check this, we plot in the inset of Fig.~\ref{compare_C} the quantity $[C(t,t_w)-C_{eq}(t-t_w)]/q_{EA}$
in order to eliminate the trivial difference of $q_{EA}$ among the different cases. Here $q_{EA}=m^2$ and 
the equilibrium squared magnetization $m^2$ has been measured numerically on the equilibrium states.
The inset shows that the scaling function
depends in a non trivial way on $a$ and this is a clear indication that the {\it superuniversality} hypothesis~\cite{cugliandolo2010, fisher1988}, 
according to which scaling functions are universal and independent on the presence/strength of the quenched
disorder, is not obeyed in the present system. 
\bfi
\begin{center}
\includegraphics*[width=0.95\columnwidth]{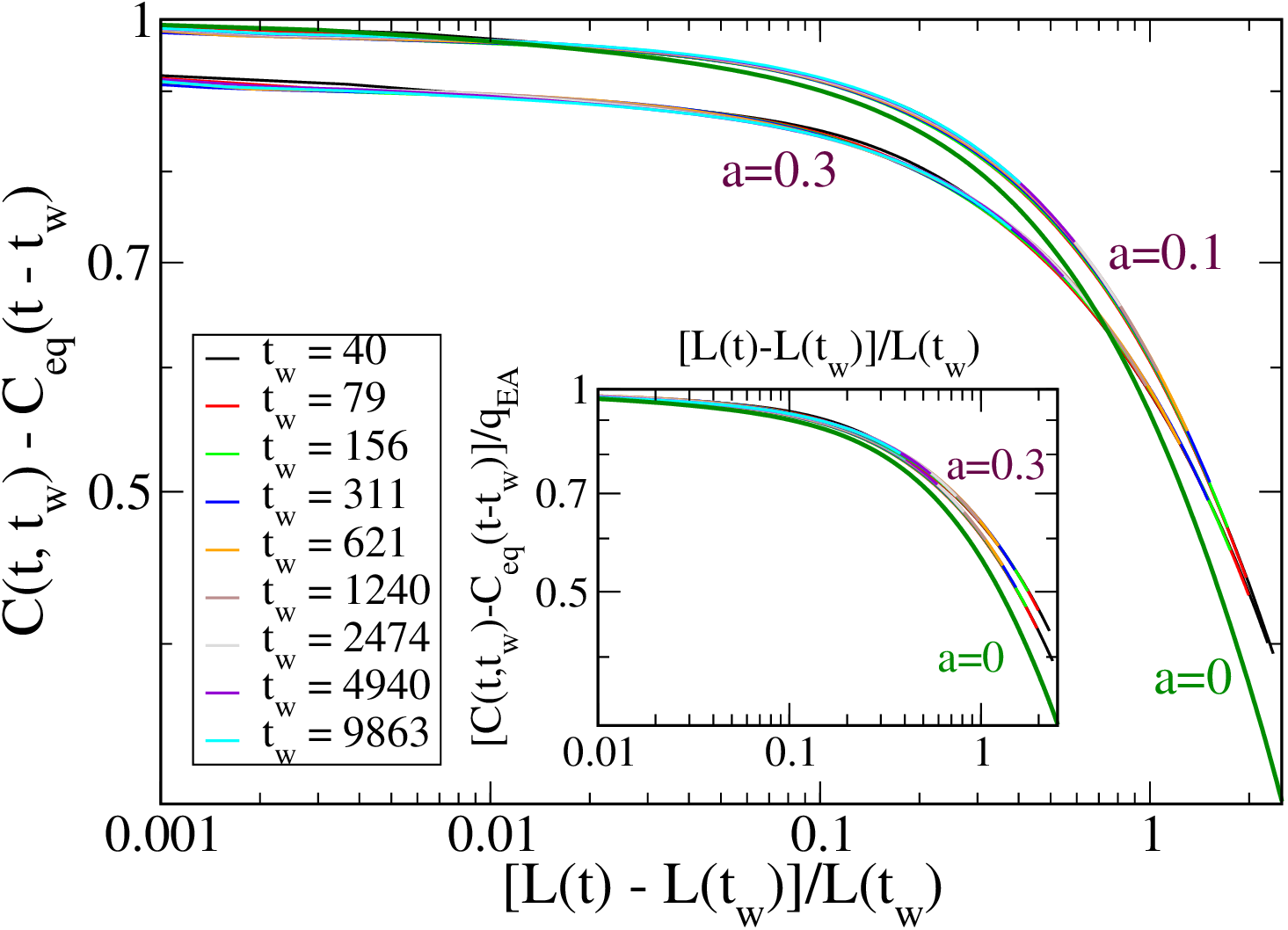}
\end{center}
\caption{$C(t,t_w)-C_{eq}(t-t_w)$ is plotted against $L(t)/L(t_w)-1$ for $a=0.1$ (rightmost set of curves, see key), and
$a=0.3$ (lower set of curves, see key). For any value of $a$ curves for different values of $t_w$ are drawn with
different colors, see key (these are difficult to distinguish because of an almost perfect data collapse). The heavy green
curve is the scaling function $c(x)$ of the pure case with $a=0$. In the inset we plot, for the same data, the quantity 
$[C(t,t_w)-C_{eq}(t-t_w)]/q_{EA}$.}
\label{compare_C}
\efi

Let us now move to the analysis of the response function. The additive scheme implies that, for the response
function, we should find data collapse by plotting $L(t_w)^\alpha [\chi (t,t_w) -\chi_{eq}(t-t_w)]$ against $L(t)/L(t_w)$ 
(see Eq.~(\ref{chiagferro})) where, in the absence
of any reference theory, the response function exponent $\alpha>0$ is considered as a fitting parameter.
This kind of plot is presented in Fig.~\ref{compare_Chi}.  
Here we see that a good collapse of the data can be achieved in the region
of large time separation (for large values of the abscissa) using values of $\alpha = 0.625, 0.2$ for 
$a=0.1,0.3$ respectively (values of $\alpha $ for different choices of $a$ are plotted in the inset). 
The value of this exponent equals the one of the low-temperature pure case $\alpha=0.625$ 
for $a=0.1$, decreases markedly upon raising $a$ and seems to vanish as $a\to a_f^-$, as it is shown in the inset of  
Fig.~\ref{compare_Chi}(b). 

Notice that the data collapse presented in Fig.~\ref{compare_Chi} is worst for smaller
values of $L(t)/L(t_w)$, but improves as $t_w$ grows larger and is always good
for the larger values of this quantity. A similar pattern is observed 
also in ferromagnetic systems without disorder~\cite{corberi2001,corberi2001b,corberi2003,corberi2003b,corberi2004b,corberi2005,corberi2005b,lippiello2006,lippiello2005,henkel2005,barrat1998,henkel2001,henkel2003,henkel2003b}. 
Let us also stress the fact that the scaling functions depend quite 
strongly on $a$, a fact that invalidates superuniversality, as already noticed studying the autocorrelation function.

\bfi
\begin{center}
\includegraphics*[width=0.47\columnwidth]{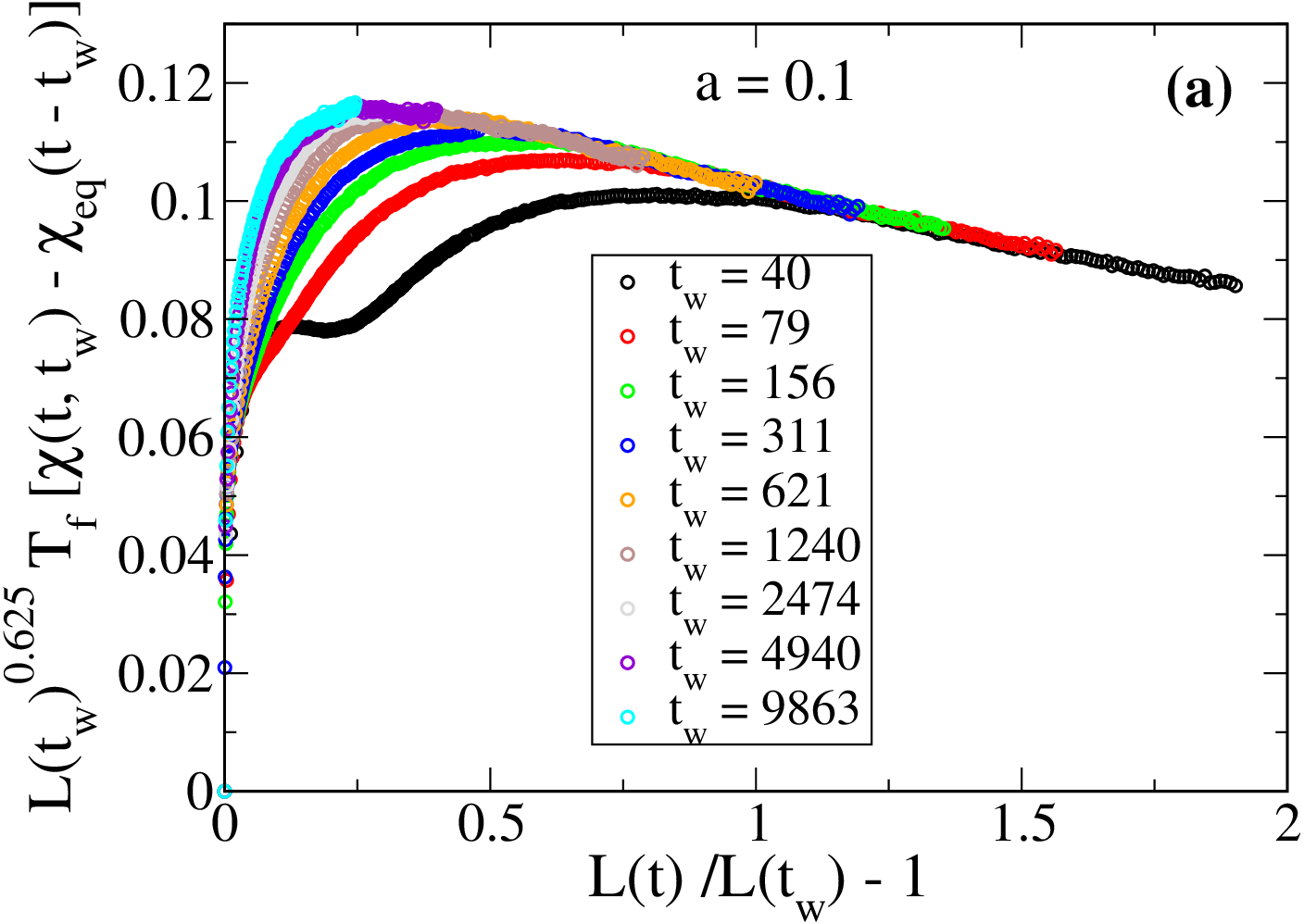}
\includegraphics*[width=0.47\columnwidth]{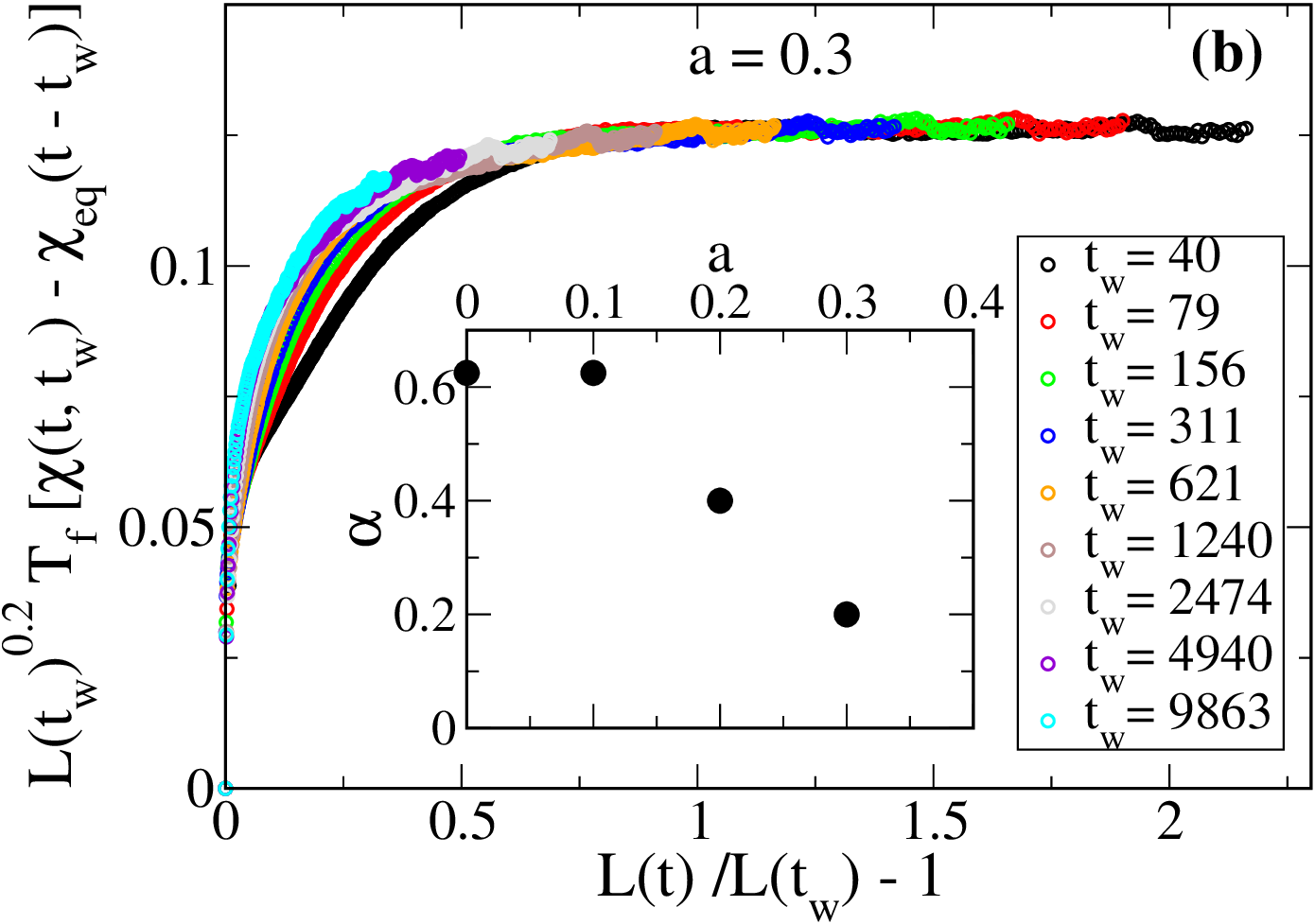}
\end{center}
\caption{$L(t_w)^\alpha T_f [\chi(t,t_w)-\chi_{eq}(t-t_w)]$ is plotted against $L(t)/L(t_w)-1$ for $a=0.1$ [left panel (a)], 
and $a=0.3$ [right panel (b)]. Curves for different values of $t_w$ are drawn with
different colors, see key. The values of $\alpha$, which are reported in the inset of the right panel, are 
$\alpha=0.625$ for $a=0.1$, $\alpha =0.4$ for $a=0.2$ and $\alpha=0.2$ for $a=0.3$.}
\label{compare_Chi}
\efi

\subsubsection{Quenches with $a= a_f$}
 
When the quench is made in a system with $a= a_f$ two main differences occur with respect to the
previous case. The first is the fact that the finite $T_f$ of our simulations corresponds to a quench into a disordered phase,
since the critical temperature $T_c(a)$ of the ferromagnetic phase goes to zero
as $a\to a_f^-$. This fact would suggest that the very asymptotic stage of the dynamics approaches the equilibrium
state rapidly. However, due to the proximity of the critical point located at zero temperature, one expects to see
slow evolution and aging in a transient preasymptotic stage. This is indeed observed in Fig.~\ref{compare_L},
where one sees that $L(t)$ keeps growing in a nearly power-law fashion at any time and there is no sign
of convergence to an equilibrium value. 

The second difference concerns the scaling properties of the two-time
quantities. Indeed, at variance with the quenches with $a<a_f$, a quench made at $T_f=0$ with $a=a_f$ is a quench
at a critical point and hence one expects a multiplicative scaling structure as the one discussed in Sec.~\ref{multiplicative}.
This is further suggested by the fact that the equilibrium magnetization vanishes at $a_f$ and hence $q_{EA}=0$ in this case.
This same structure should characterize the two-time quantities also for quenches to finite temperatures,
as the one we are studying numerically, provided they are sufficiently low in order to have a long-lasting aging stage. This is the case we consider here
since for a quench right
at $T_f=0$ the system gets trapped in metastable states.

In Fig.~\ref{autocdivceqL_T075_a04} we plot the quantity $C(t,t_w)/C_{eq}(t-t_w)$ against $L(t)/L(t_w)$ which,
according to Eqs.~(\ref{Cmult},\ref{scalCcrit}) should amount to $C_{ag}$ and provide data collapse of the curves
with different $t_w$. Indeed this is what one observes with great precision. This confirms that the multiplicative 
structure of a critical point is present. Let us add that the additive structure is definitely ruled out in this case also
by the fact that $C_{eq}(t,t_w)>C(t,t_w)$ for any $t>t_w$, so that, if an additive scheme would apply, one should have
$C_{ag}=C-C_{eq}<0$, which in unphysical. 

For completeness, let us briefly discuss the behavior of $C_{eq}$, which is plotted in the inset of Fig.~\ref{autocdivceqL_T075_a04}.
It decays approximately  as in Eq.~(\ref{Ceqmult}) with a very small exponent
$B\simeq 0.005$. From the data of Fig.~\ref{compare_L} we see that, for sufficiently long times,  
$L(t)$ grows approximately in an algebraic way $L(t)\sim t^{1/z}$, with $z\simeq 5$
(this is the dashed indigo line in the figure). Hence, for the exponent $\gamma $ defined in Eq.~(\ref{distance}),
we find $\gamma \simeq 0.025$, a fact that we will use soon.
\bfi
\begin{center}
\includegraphics*[width=0.95\columnwidth]{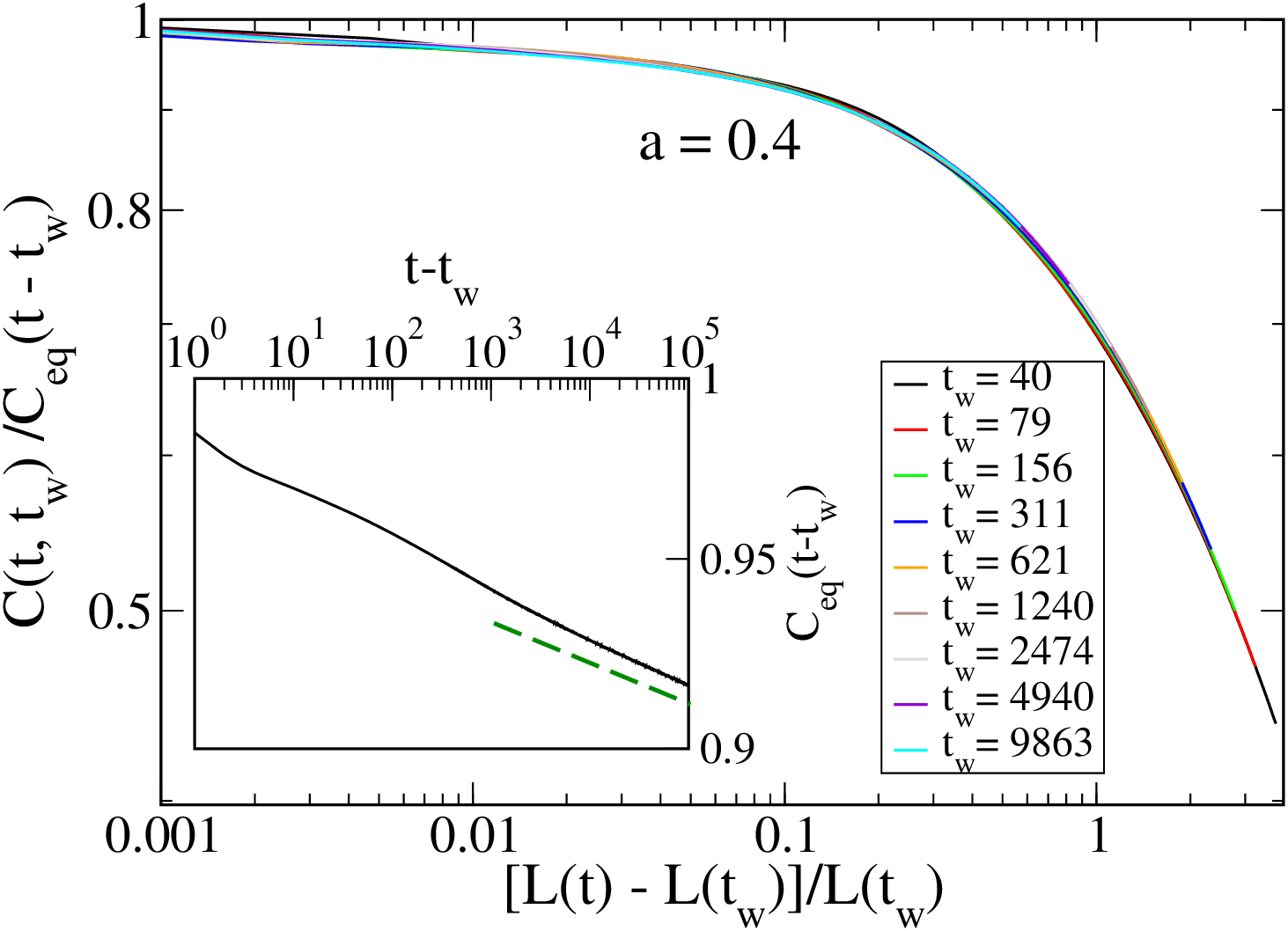}
\end{center}
\caption{$C(t,t_w)/C_{eq}(t-t_w)$ is plotted against $L(t)/L(t_w)-1$ for $a=0.4$.
Curves for different values of $t_w$ are drawn with
different colors, see key. In the inset the equilibrium correlation $C_{eq}(t-t_w)$ is plotted against
$t-t_w$. The green dashed line is the algebraic form $x^{-0.005}$.}
\label{autocdivceqL_T075_a04}
\efi

We turn now to the discussion of the response function.  According to the multiplicative scheme,
one should find data collapse by plotting $L(t_w)^\gamma [1-T_f\chi (t,t_w)]$ against $L(t)/L(t_w)$, as expressed
by Eq.~(\ref{distance}).
The value of $\gamma \simeq 0.025$ has been estimated above from the properties of the autocorrelation function.
We see in Fig.~\ref{chi_a04_risc} that this value produces a good collapse of our data. Some residual oscillations, which are present
particularly in the curve with smaller $t_w$, spoil somewhat the superposition, but these oscillation tend to decrease
as $t_w$ is taken larger and the collapse for the corresponding curves improves progressively.

It should be noted that the structure found in this quench cannot be framed among 
the scaling paradigms discussed in Sec.~\ref{scaling} for clean ferromagnetic systems, as we explain below.
Since $T_c=0$ for
$a=a_f$, with this value of $a$ the model can be interpreted as being at $d_L$.
As discussed in Sec.~\ref{scaling}, in clean systems this would imply $\alpha=0$.
We see in the inset of Fig.~\ref{compare_Chi}, indeed, that the behavior of $\alpha $ as $a$ is varied
is consistent with the vanishing of this exponent as $a\to a_f$.
In a clean system at $d=d_L$, however, one has a finite value of $q_{EA}$, since the model is
fully ordered at $T=0$, which implies as additive scheme. In this case, instead,
the point $(a=a_f, T=0)$ is associated to a vanishing ferromagnetic order parameter, since it is the frontier
with the paramagnetic region. Similarly, it is plausible that the spin-glass order parameter, which is finite
at $T=0$ for $a>a_f$, also vanishes as $a\to a_f^+$. 
From this point of view, then, the multiplicative structure that
we find could be legitimate, since any order parameter vanishes in this critical point.

As a final remark, let us also mention that in clean magnetic systems with a scalar order parameter at $d=d_L$, as the Ising model
in $d=1$, the equilibrium parts of both $C$ and $\chi$ vanish identically. This is not true in the present model at 
$a=a_f$, as we have shown. In a sense, the situation is reminiscent of what one has in a clean system with a vectorial order
parameter, since in that case the response function is finite even at $T=0$ due to the existence of the Goldstone modes.
Possibly, the presence of soft modes is the origin of finite $C_{eq}$ and $\chi _{eq}$ also in the present model. These soft modes
could arise as due to the peculiar character of interfaces, whose rearrangements might be enhanced by frustration as compared
to what occurs in the ferromagnetic region.

 \bfi
\begin{center}
\includegraphics*[width=0.95\columnwidth]{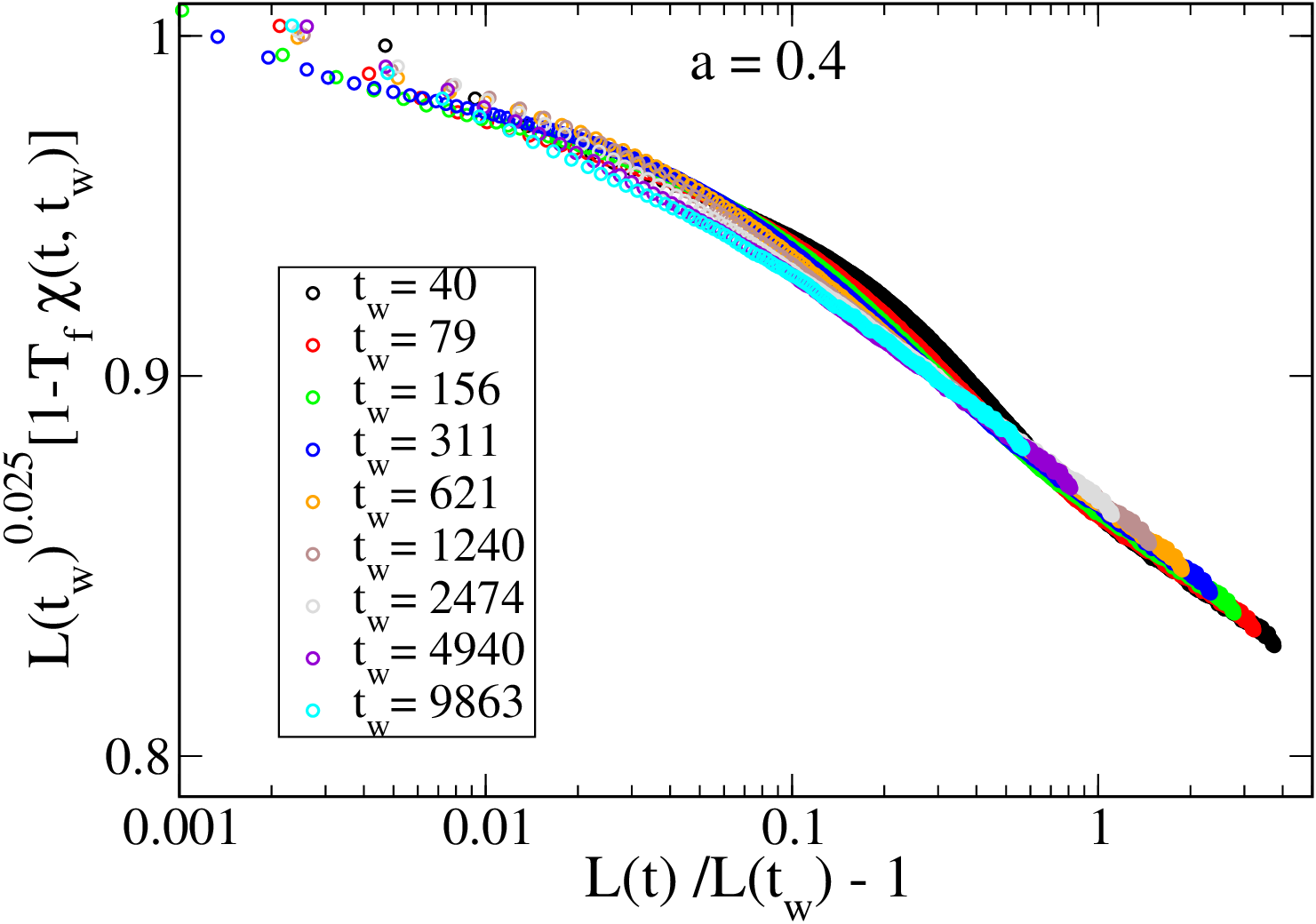}
\end{center}
\caption{$L(t_w)^{0.025}[1-T_f\chi(t,t_w)]$ is plotted against $L(t)/L(t_w)-1$ for $a=0.4$.
Curves for different values of $t_w$ are drawn with
different colors (see key).}
\label{chi_a04_risc}
\efi

\subsubsection{Quenches with $a> a_f$}

Even if the state of the system is disordered for any finite temperature when $a>a_f$ the steady growth of $L(t)$
observed in Fig.~\ref{compare_L}
signals that, in the range of times accessed in the simulations, the system is far from equilibration
and the kinetics is slowed down due to the proximity of a critical region. Indeed we know that not only a critical point
exists at $(a=a_f, T=0)$, but also the whole region $(a_f<a<a_a, T=0)$ is presumably interested by
a spin-glass phase~\cite{corberi2017}. For this reason we expect to detect scaling properties for the two-time 
quantities. However, for this type of quenches, it is not obvious which scaling structure could emerge.
Indeed, the system could feel the critical point at $(a=a_f,T=0)$, in which case one would expect 
basically the same multiplicative structure observed in the quench at $a=a_f$ and with the same exponents.
On the other hand, it is also possible that the influence of the spin-glass phase determines the behavior
of the system. In this case, since in the spin-glass phase one has $q_{EA}$, one could expect an additive
structure to be appropriate. 

Let us discuss the numerical data. 
First of all, as for the quench with $a>a_f$, we find that $C_{eq}(t-t_w)>C(t,t_w)$ for any $t>t_w$,
a fact that rules out the additive scheme. We show in Fig.~\ref{compare_C_para} that, indeed, the 
multiplicative structure is very well verified. We mention that a similar multiplicative form for
the autocorrelation was also found in the $3d$ spin-glass phase~\cite{berthier2002,manssen2015},
although the matter is debated since different interpretations~\cite{jaubert2007} may support either
an additive structure. In our case, instead, this is definitely ruled out.
Notice that also in this case the two curves depend on the amount of disorder $a$.
 
The equilibrium part $C_{eq}$ of the autocorrelation (inset of Fig.~\ref{compare_C_para}) has a power law decay 
 as in Eq.~(\ref{Ceqmult}), with an exponent $B\simeq 0.0162$.   
 We stress that a power law behavior of $C_{eq}$ is usually
 found in spin-glass phases (for $d\ge 3$)~\cite{berthier2002,manssen2015}.  
 Finally, let us remark that
 the oscillations of $C_{eq}$ in the case $a=0.7$ does not allow a clear statement about the behavior
 of this function. 

 \bfi
\begin{center}
\includegraphics*[width=0.95\columnwidth]{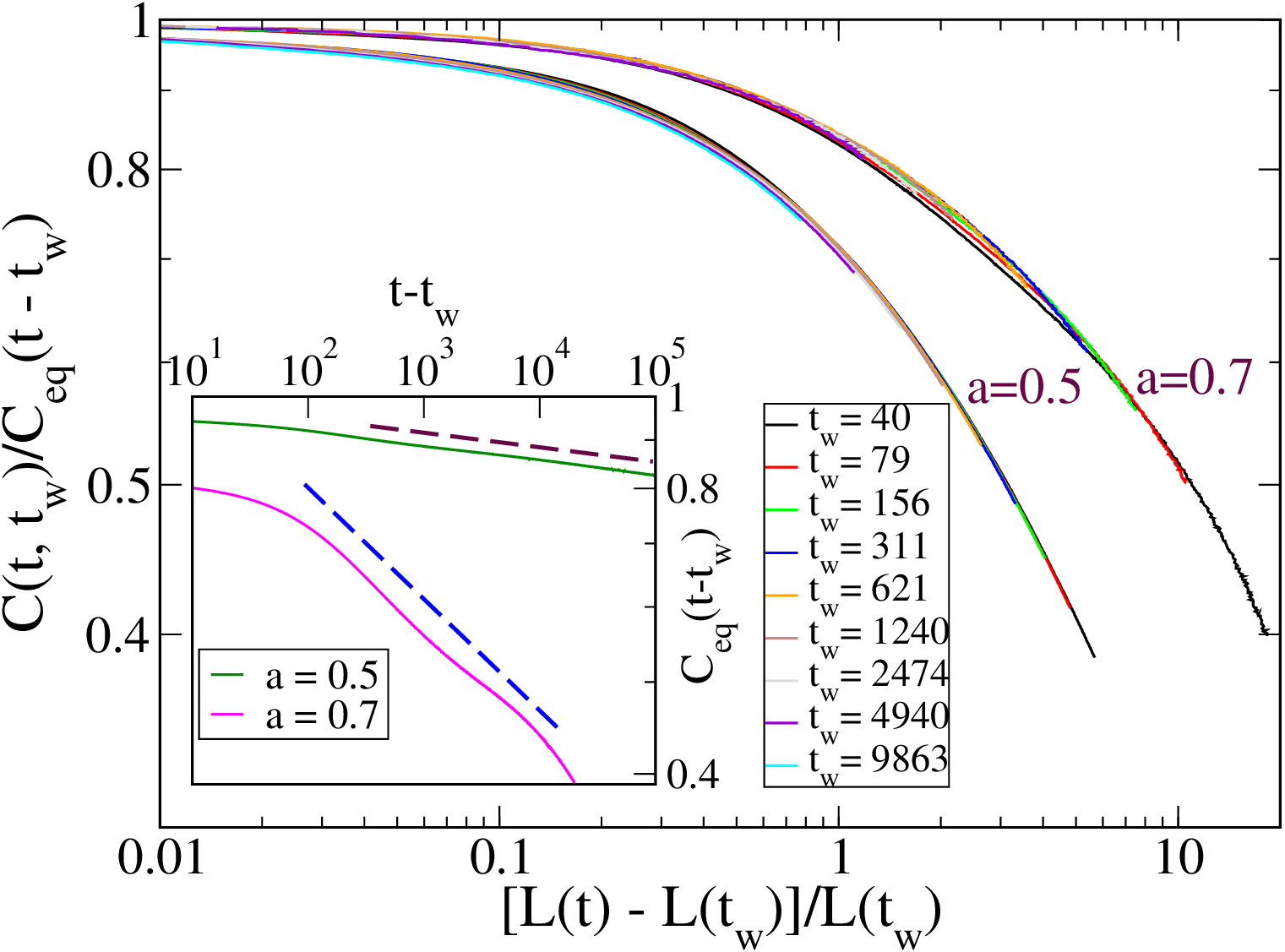}
\end{center}
\caption{$C(t,t_w)/C_{eq}(t-t_w)]$ is plotted against $L(t)/L(t_w)-1$ for $a=0.5$ (lower set of curves) and
$a=0.7$ (upper set of curves).
Curves for different values of $t_w$ are drawn with
different colors, see key. In the inset the quantity $C_{eq}(t-t_w)$ is plotted for $a=0.5$ and $a=0.7$. The dashed
maroon line is the behavior $x^{-0.0162}$ and the dashed blue is $x^{-0.12}$.}
\label{compare_C_para}
\efi

Let us now discuss the behavior of the response function. In Fig.~\ref{chiL_T075_a05_resc}
we plot the quantity $L(t_w)^\gamma [1-T_f\chi(t,t_w)]$ against $L(t)/L(t_w)-1$ which, recalling the discussion
above and Eq.~(\ref{distance}), should result in a data collapse using $\gamma=Bz$.
For $a=0.5$, from the data of Fig.~\ref{compare_L} we see that for sufficiently long times  
$L(t)$ grows approximately in an algebraic way $L(t)\sim t^{1/z}$, with $z\simeq 4.07$ (this is the brown dashed line)
and hence $\gamma \simeq 0.066$. We can see in Fig.~\ref{chiL_T075_a05_resc} that a good
data collapse for the response function is obtained for a somewhat larger value $\gamma \simeq 0.07$. Given the noisy character of the problem and the 
possible presence of preasymptotic corrections we consider this value compatible with the general scaling framework.

For $a=0.7$ we get a good superposition of the curves with different $t_w$ with $\gamma =0.3$. Notice that, both for $a=0.5$ and $a=0.7$, the collapse 
starts to be good earlier for large time differences (large $L(t)/L(t_w)$) and is worse for smaller time.
This feature, however, is much more enhanced for $a=0.7$. Nevertheless, also for $a=0.7$ the
superposition is satisfactory for the curves with larger $t_w$ basically in the whole range of $L(t)/L(t_w)$.
Using the growth exponent value $z\simeq 2.56$ obtained from the curve of $L(t)$ with $a=0.7$ in Fig.~\ref{compare_L},
together with $\gamma =0.3$, we get $B=\gamma /z\simeq 0.12$. In the inset of Fig.~\ref{compare_C_para} we see that, indeed, 
this value of $B$ is consistent with the decay of $C_{eq}$, despite the presence of oscillations does not allow us to reach
a definite conclusion.

\bfi
\begin{center}
\includegraphics*[width=0.95\columnwidth]{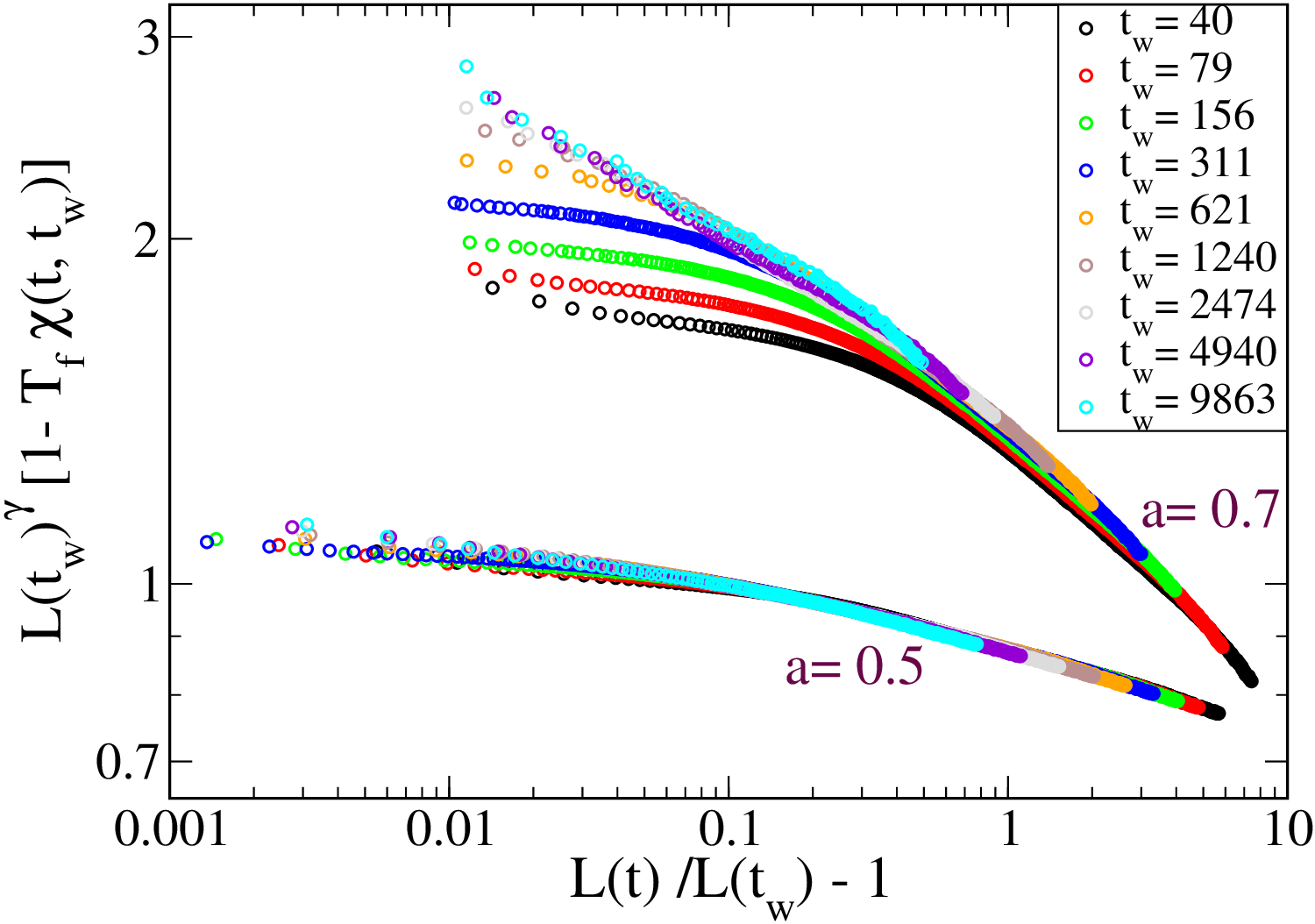}
\end{center}
\caption{$L(t_w)^\gamma [1-T_f\chi(t,t_w)]$ is plotted against $L(t)/L(t_w)-1$ for $a=0.5$ (lower set of curves)
and $a=0.7$ (upper set of curves).
Curves for different values of $t_w$ are drawn with
different colors, see key. The value of $\gamma $ is $\gamma=0.07$ for $a=0.5$ and $\gamma=0.3$ for $a=0.7$}
\label{chiL_T075_a05_resc}
\efi

\subsection{Fluctuation dissipation plot}

In equilibrium the response function can be written in terms of the autocorrelation function using the
fluctuation-dissipation theorem. Indeed, since both $\chi_{eq}$ and $C_{eq}$ depend only
on $t-t_w$ a parametric form $\chi_{eq}(t-t_w)=\widetilde \chi (C)$, with $T\widetilde \chi(C)=1-C$
(we stick here to spin systems) of the response in term of the autocorrelation can be arrived at. 
Out of equilibrium, when a non-trivial dependence on both the two times occurs, such a parametrization
is not, in principle possible. However, since $C$ is usually a monotonic function, one can eliminate
one of the two times, say $t$, from $\chi(t,t_w)$ in favor of $C$, thus obtaining $\chi(t,t_w)=\widehat \chi(C,t_w)$,
 and look at the parametric representation
(or fluctuation dissipation plot)
\be
\widetilde \chi(C)=\lim _{t_w\to \infty}\widehat \chi(C,t_w),
\ee
if this limit exists. This relation is of great interest since it represents a bridge between the non-equilibrium
dynamic properties, embodied by $\chi$ 	and $C$, and the static equilibrium ones, represented by the 
overlap probability distribution $P(q)$~\cite{franz98}.

With the scaling forms of the two-time quantities discussed in section~(\ref{scaling}),
for the additive (with $d>d_L$) and multiplicative cases in non-disordered systems 
one finds~\cite{corberi2004,corberi2007}
the behavior that is schematically shown in Fig.~\ref{schematicfdt}.
In this figure not only we plot the limiting form $\widetilde \chi (C)$ but also the
approach of $\widehat \chi (C,t_w)$ to $\widetilde \chi (C)$ as $t_w$ is progressively increased.

With the additive form one has the broken line
\be
T_f \widetilde \chi (C)= \left \{ \begin{array}{ll}
1-q_{EA}, & \quad\mbox{for }C<q_{EA} \\
1-C, & \quad\mbox{for }C\ge q_{EA}.
\end{array} \right .
\label{brokenline}
\ee
In this case the approach of $\widehat \chi(C,t_w)$ to $\widetilde \chi (C)$ as $t_w$ increases is quite rapid for $C>q_{EA}$, while
it is much slower and occurs from above in the region $C<q_{EA}$. It can be easily realized that the overshoot of $\widehat \chi$ with respect
to $\widetilde \chi$ in this region is due to the non-equilibrium contribution $\chi _{ag}$ whose asymptotic vanishing is regulated by the
exponent $\alpha$, see Eq.~(\ref{chiagferro}).

With the multiplicative structure, instead, the asymptotic form is the line $T_f\widetilde \chi (C)=1-C$, as in equilibrium.
The approach of $\widehat \chi(C,t_w)$ to $\widetilde \chi (C)$ as $t_w$ is increased is from below and convergence
starts from larger values of $C$ (Fig.~\ref{schematicfdt}(b)) and then progressively takes place at lower and lower values of $C$.

\bfi
\begin{center}
\includegraphics*[width=0.47\columnwidth]{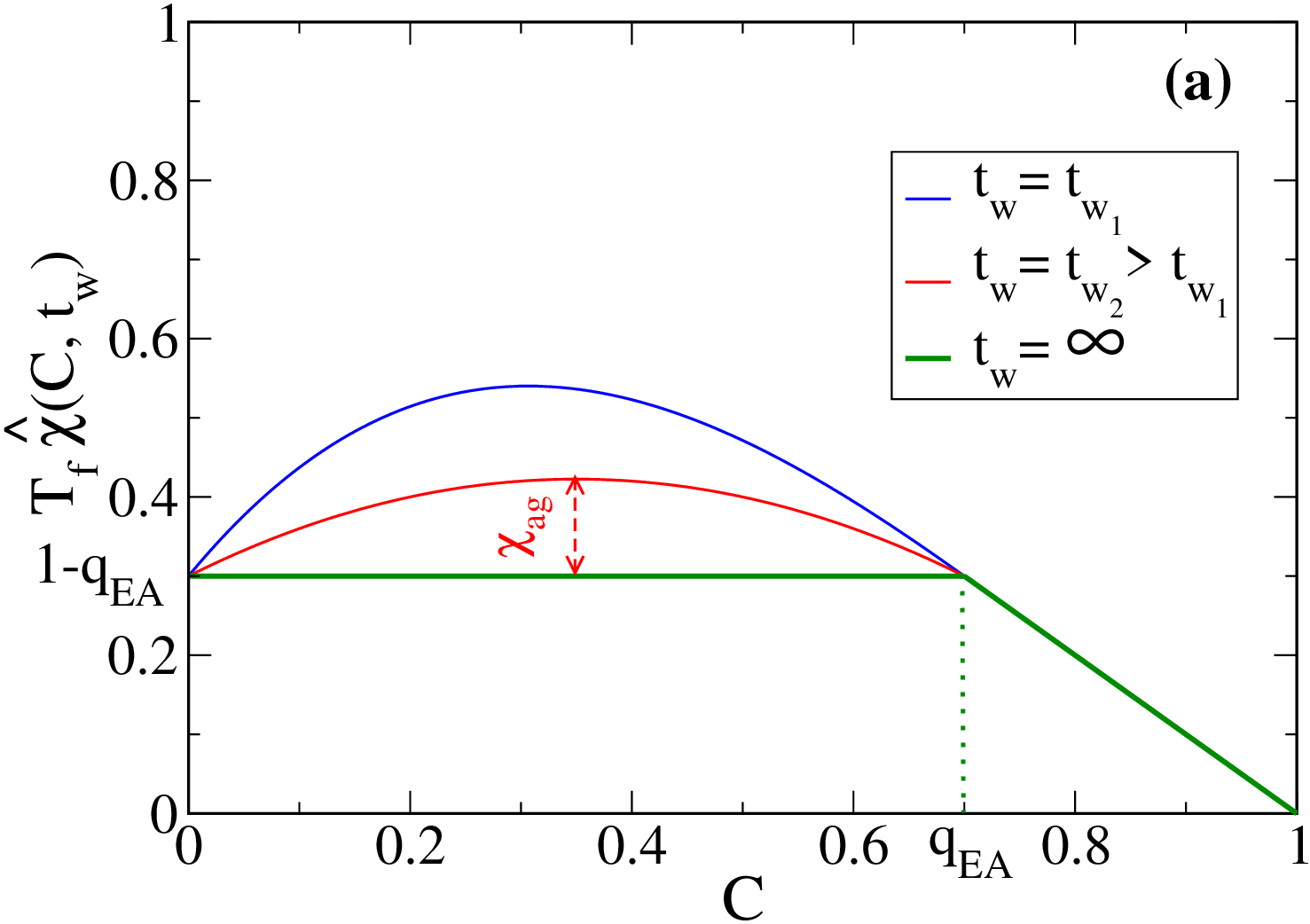}
\includegraphics*[width=0.47\columnwidth]{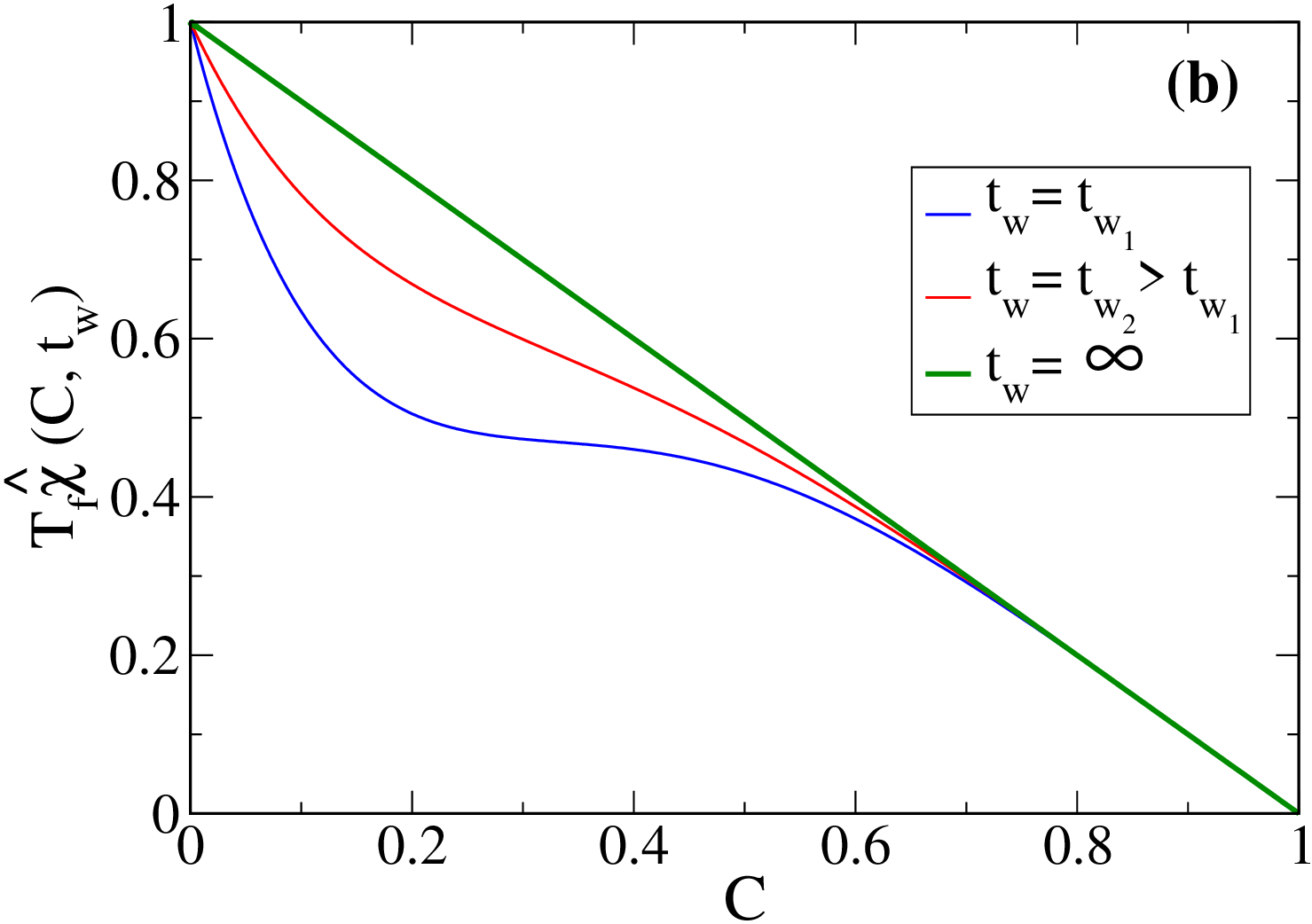}
\end{center}
\caption{Schematic representation of the approach of the curve $T_f\widehat \chi (C,t_w)$ to the asymptotic
form $T_f\widetilde \chi(C)$ for a system with an additive scaling [left panel (a), see Sec.~\ref{additive}], and a multiplicative scaling [right panel (b), see Sec.~\ref{multiplicative}].
Curves for different values of $t_w$ are drawn with
different colors, see key. The bold green curve with $t_w=\infty$ represents $T_f\widetilde \chi (C)$.}
\label{schematicfdt}
\efi

Let us now see how our numerical data behave. 
Examples of parametric plots for $a<a_f$ are shown in Fig.~\ref{compare_autochi_ferro}.
In this figure the asymptotic form~(\ref{brokenline}) has been drawn by taking $q_{EA}=m^2$ and 
computing the equilibrium squared magnetization $m^2$ numerically on the equilibrium states.
Notice that $1-q_{EA}\simeq 0$ for $a=0.1$ since the equilibrium magnetization is $m\simeq 1$.

Regarding the approach of $\widehat \chi(C,t_w)$ to $\widetilde \chi (C)$, one observes a pattern 
qualitatively similar to the one shown in Fig.~\ref{schematicfdt}(a).
Of course, the convergence is slow due to the limited range of $t_w$. Furthermore, since the
overshoot of $\widehat \chi$ with respect to $\widetilde \chi$ is due
to $\chi _{ag}$, it disappears slower the smaller $\alpha $ is, see Eq.~(\ref{chiagferro}) and discussion above.
Since $\alpha$ gets smaller upon raising $a$, this determines a slower convergence of the curves
with $a=0.3$ as compared to those with $a=0.1$. Besides that, also the speed of growth
of $L(t)$, which is a bit larger for $a=0.1$ than for $a=0.3$, plays a role in fostering the convergence of the curves.
Notice also the anomalous feature (with respect to clean systems) of a slow convergence (from below) of the curves also
for $C>q_{EA}$, a fact that is possibly due to the slow character of the equilibrium states themselves.

\bfi
\begin{center}
\includegraphics*[width=0.95\columnwidth]{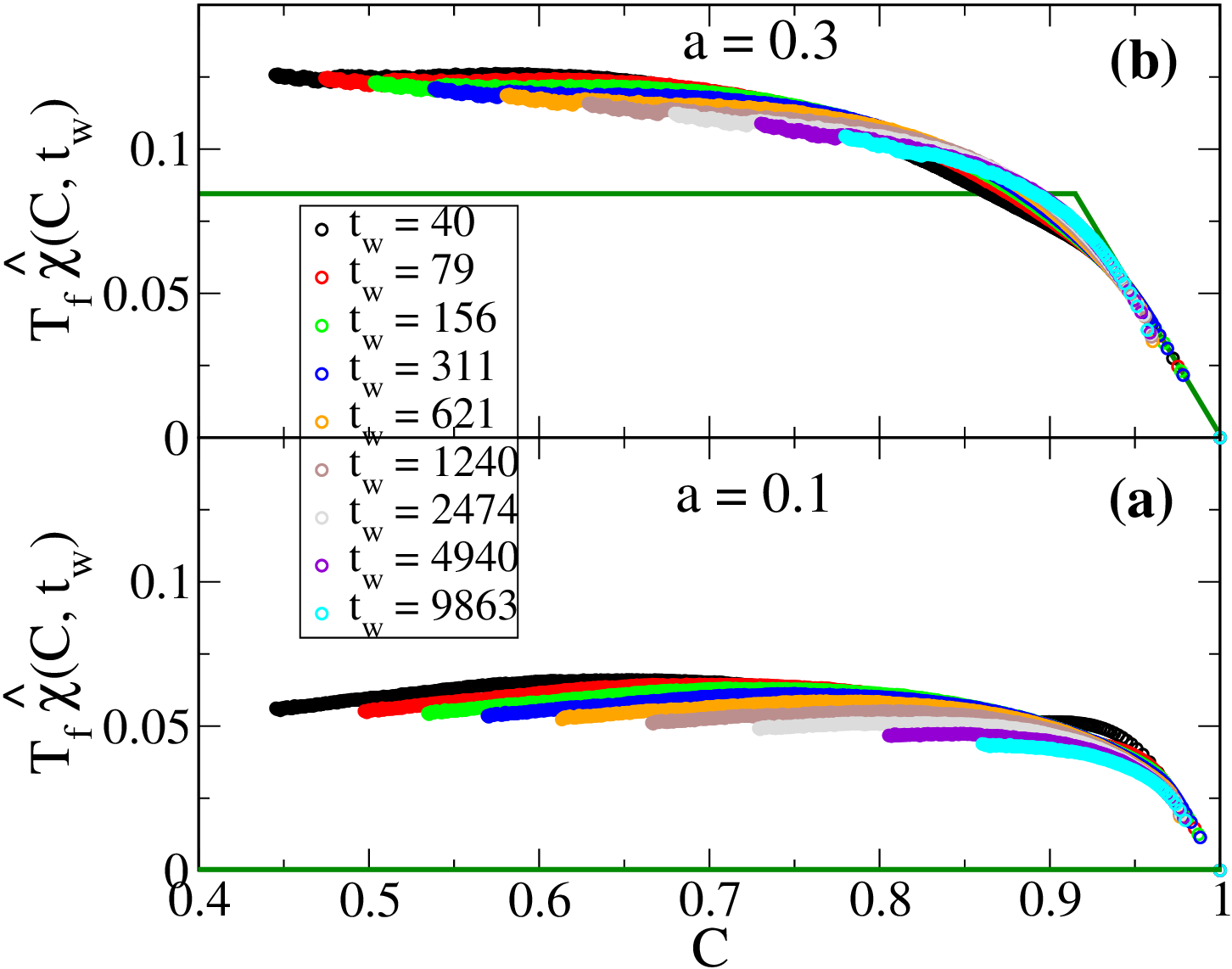}
\end{center}
\caption{$T_f\widehat \chi(C,t_w)$ is plotted against $C$ for $a=0.1$ [lower panel (a)]
and $a=0.3$ [upper panel (b)].
Curves for different values of $t_w$ are drawn with
different colors, see key. The bold green lines are the asymptotic forms $T_f \widetilde \chi (C)$.}
\label{compare_autochi_ferro}
\efi

According to our previous results, a multiplicative scheme applies in the whole region with $a\ge a_f$.
Hence, in this region we expect to see a parametric plot qualitatively similar to the Fig.~\ref{schematicfdt}(b). Data from the simulations are presented in Fig.~\ref{compare_autochi_mult},
confirming the expected behavior. In particular, for $a=0.5$ the convergence towards $T_f\widetilde \chi(C)=1-C$
is rather slow, due to the very small value of the exponent $\gamma$ ($\gamma \simeq 0.07$) and also to the
slow growth of $L(t)$.
For $a=0.7$, $\gamma $ takes the larger value $\gamma \simeq 0.3$ and $L(t)$ grows much faster, and this fact
greatly speeds up the convergence of the curves. Indeed, we see that at the  largest values of $t_w$ the
curve $\widetilde \chi (C)$ is almost attained.

\bfi
\begin{center}
\includegraphics*[width=0.95\columnwidth]{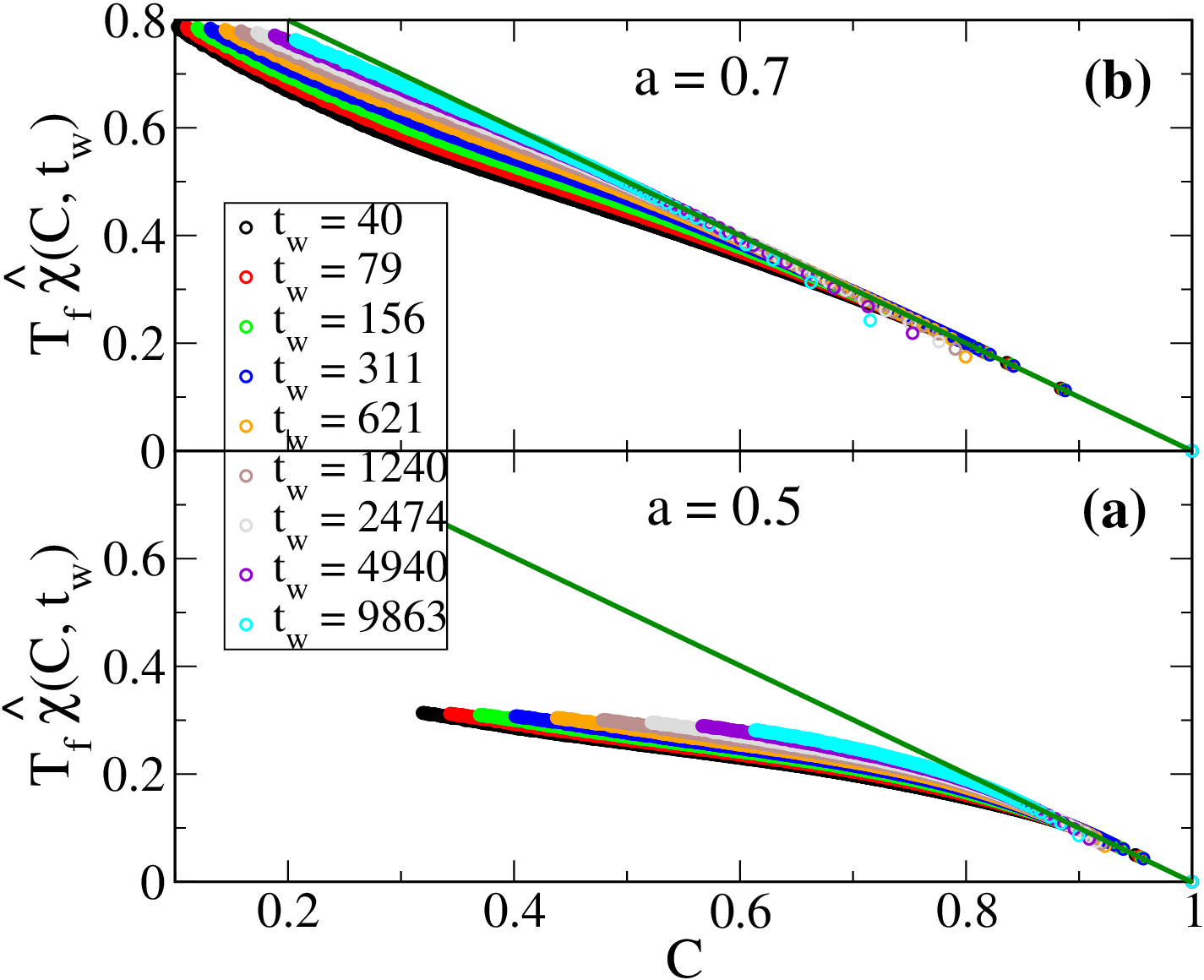}
\end{center}
\caption{$T_f\widehat \chi(C,t_w)$ is plotted against $C$ for $a=0.5$ [lower panel (a)]
and $a=0.7$ [upper panel (b)].
Curves for different values of $t_w$ are drawn with
different colors, see key. The bold green lines are the asymptotic form $T_f \widetilde \chi (C)$.}
\label{compare_autochi_mult}
\efi

\section{Conclusions} \label{concl}

The aim of this paper was to study how the scaling properties of two-time quantities,
specifically the autocorrelation function and the associated response function, are affected by the presence
of disorder and frustration in the kinetics of a magnetic system after a quench to a low temperature.
In order to do that, we have studied numerically the model discussed
in~\cite{corberi2017}, which amounts to an Ising system in two dimensions where a fraction $a$ of
couplings take a negative value, while the remaining ones are ferromagnetic. Varying $a$ this model
interpolates between a clean ferromagnet at $a=0$, a disordered ferromagnet for $0<a<a_f$, and a
paramagnet with a zero-temperature spin-glass phase for $a>a_f$.

Being two-dimensional, the model has the
advantage of a relatively fast determination of the equilibrium states at low temperature in rather
large systems, a fact that allows one to compute the equilibrium behavior of the two-time quantities
with good accuracy. This has been utilized to address the issue of how equilibrium and aging degrees of
freedom contribute to the kinetic evolution and how they combine to form the two-time quantities.

Our results show that this occurs quite differently in the different phases of the model.
In particular, in the whole ferromagnetic region, for $0\le a\le a_f$, one observes the same additive structure 
$C=C_{eq}+C_{ag}$ (and similarly for the response function) where the equilibrium and the aging parts sum up
to form the complete correlation and response. This was expected and is consistent with the existence of
a finite order parameter.
Though the additive property  applies to the entire ferromagnetic
phase, the actual behavior of the two-time quantities turns out to be strongly dependent on
the amount $a$ of disorder. The response function exponent $\alpha$, in particular, and the scaling functions,
depend on $a$. This shows quite clearly that the property of superuniversality is not obeyed.

The fact that the response function exponent $\alpha $ vanishes in the limit $a\to a_f^-$ is of interest.
Indeed this is what happens in clean magnets at the lower critical dimensionality and, in that context,
it is interpreted as
due to the fact that interfaces are free to move without experiencing any restoring force.
For instance, Ising interfaces in $d=1$ are pointlike random walkers.
In the present case, one can provide a similar interpretation.
Indeed, the model is at the lower critical dimension when $a=a_f$, since the critical temperature for ferromagnetism
vanishes, see Fig.~\ref{fig_phase_diagr}. Furthermore, it is conceivable that, due to the large amount of
negative bonds along interfaces when $a=a_f$, these become soft objects whose displacement can occur rather freely
as opposed to a clean (or weakly disordered) two-dimensional magnets where motion is driven by surface tension and
curvature.

The additive structure breaks down at the critical point at $a=a_f$, and one sees a multiplicative one
with $C=C_{eq}\cdot C_{ag}$. This might be considered consistent with what we know in clean magnets,
since in that case a multiplicative structure emerges when the system is quenched to a critical point
with a vanishing order parameter.

The same multiplicative structure is found when the system is quenched in the paramagnetic region
with $a>a_f$. It must be observed that, although our quenches are done to finite temperatures and hence
in the disordered phase of the model, in the range of times accessed in the simulations the system does not
show any sign of equilibration. This can be ascribed to the zero-temperature spin-glass phase
extending its influence to the preasymptotic evolution of the model. According to this interpretation,
since in the spin-glass phase there is a non-vanishing order parameter, one should expect to find
an additive structure for the two-time quantities. Instead, we have a clear indication of a multiplicative
scheme, as already observed in other spin-glass systems~\cite{berthier2002,manssen2015},
whose meaning remains to be clarified. 

As a final remark, let us comment on the fact that the results of this paper, besides the interest
in addressing general properties of the non-equilibrium kinetics of slowly relaxing systems, can
help elucidating the structure of the phase-diagram of frustrated systems. In particular, the recognition
of a different scaling paradigm, additive vs multiplicative, might enable one to distinguish between
different phases, a fact that could turn out to be relevant and useful in the controversial field of
frustrated systems. In this respect, the investigation of the properties of
two-time quantities
along the lines followed here
in the three dimensional case represents an interesting research project for the future. 

\vspace{1cm}

\noindent
{\bf Acknowledgements}

Numerical results were obtained using the High-Performance Computing facility
at IUAC, New Delhi (http://www.iuac.res.in/labs/hpcl/index.html). 
F.C. acknowledges financial support by MIUR project PRIN2015K7KK8L.

\end{document}